\documentclass[letter,twocolumn,showpacs,preprintnumbers,pra,byrevtex]{revtex4}
\usepackage{graphics}% Include figure files
\usepackage{amsmath}
\usepackage{amssymb}
\usepackage{bm}% bold math
\begin{document}

%\preprint{Submitted to Physical Review A on 12 April 2003}

\title{``Quality'' of a Which-Way Detector}% Force line breaks with \\

\author{Jes\'us Mart\'{\i}nez-Linares}
\thanks{corresponding author}
 %\altaffiliation[Also at ]{}
%\author{}
\affiliation{Facultad de Ingenier\'{i}a en Tecnolog\'{i}a de la Madera. \\Edificio D. Ciudad Universitaria. Universidad Michoacana de San Nicol\'{a}s
de Hidalgo.\\ 58060 Morelia, Michoac\'{a}n, M\'{e}xico.}

\author{David A. Harmin}
 \homepage{http://www.pa.uky.edu/~mike/amo_har.html}
\affiliation{
Department of Physics and Astronomy, University of Kentucky, \\
Lexington, KY 40506--0055, USA}

%\date{\today}
%\date{Received \phantom{31 December 1999}}

\begin{abstract}
We introduce a measure~$\mathcal{Q}$ of the ``quality'' of a
quantum which-way detector, which characterizes its intrinsic
ability to extract which-way
information in an asymmetric two-way interferometer. The ``quality''~$\mathcal{Q%
}$ allows one to separate the contribution to the distinguishability of the
ways arising from the quantum properties of the detector from the
contribution stemming from \textit{a~priori} which-way knowledge available to
the experimenter, which can be quantified by a predictability parameter ~$%
\mathcal{P}$. We provide an inequality relating these two sources
of which-way information to the value of the fringe visibility
displayed by the interferometer. We show that this inequality is
an expression of duality, allowing one to trace the loss of
coherence to the two reservoirs of which-way information
represented by~$\mathcal{Q}$ and~$\mathcal{P}$. Finally, we
illustrate the formalism with the use of a quantum logic gate: the
Symmetric Quanton-Detecton System (SQDS). The SQDS can be regarded
as two qubits trying to acquire which way information about each
other. The SQDS  will provide an illustrating example of the
reciprocal effects induced by duality between system and which-way
detector.
\end{abstract}

\pacs{42.50.-p  03.67.-a 42.50.Xa}
%42.50.-p Quantum Optics
%03.67.-a Quantum Information
%42.50.Xa Optical test of quantum theory
\maketitle

\section{\label{sec:Introduction}Introduction}

\textit{The observation of an interference pattern and the
acquisition of which-way information are mutually exclusive.} This
statement, which is often quoted as a definition of the duality
principle~\cite{Englert96}, has driven the debate on the
fundamentals of Quantum Mechanics since its
foundation~\cite{Feynman65, Scully91}. A very elegant approach has
been developed by Englert~\cite{Englert96}, which allows one to
quantify the notion of wave-particle duality of a quantum system
(the ``Quanton'') in a two-way interferometer. He derives
\textit{an inequality concerning duality}, according to which the
fringe visibility~$\mathcal{V}$ displayed at the output port of
the interferometer sets an absolute upper bound on the amount of
which-way information~$\mathcal{D}$ that is potentially stored in
a generic which-way detector (WWD)\@. Here $\mathcal{D}$ is the
{\em distinguishability} of the two ways defined
in~\cite{Englert96}. The inequality reads
\begin{equation}
\mathcal{D}^2 +\mathcal{V}^2 \le 1, \label{1}
\end{equation}
encoding the extent to which partial which-way information and
partial fringe visibility are compatible. In particular, the
extreme situations characterized by perfect fringe visibility
($\mathcal{V}=1$) or full which-way information ($\mathcal{D}=1$)
are mutually exclusive, so the bound in~(\ref{1}) can be
interpreted as an expression of duality. Inequalities of this type
involving Duality have attracted a great interest, both
theoretically \cite{Englert2000, Bjork98} and experimentally
\cite{Schwindt99, DurrNature98, DurrPRL98, Durr2000}.

Nevertheless, two different sources of which-way information are
represented in $\mathcal{D}$. One is the \textit{predictability}
of the ways $\mathcal{P}$, i.e., the \textit{a~priori} which-way
knowledge that the experimenter has about the ways stemming from
the preparation of the beam splitter and the initial state of the
Quanton. This inherent lopsidedness of the system is simply a bias
for one or the other way built into the initial state, as in a
``loaded'' coin toss. The second source of which-way information
contained in $\mathcal{D}$ is purely quantum mechanical, stemming
from the WWD's ability to correlate the two ways with two or more
of its own final states, leading to the ``storage'' of some
which-way information in the detector.

It is the purpose of this paper to investigate the case of
interferometers characterized by $\mathcal{P}\ne0 $ in order to
sort out how much of the loss of fringe visibility originates in
the predictability and how much results from the inherent quantum
properties of the detector. To achieve this goal we introduce a
measure~$\mathcal{Q}$ of the ``quality'' of the detector, which,
roughly speaking, measures how good the WWD is. The
parameter~$\mathcal{Q}$ characterizes the intrinsic ability of the
detector to extract which-way information via quantum
correlations. We then derive an inequality that treats the three
quantities $\mathcal{P}$, $\mathcal{Q}$, and~$\mathcal{V}$ on an
equal footing. This formalism allows one to trace the loss of
coherence in asymmetric interferometers quantitatively to the two
reservoirs of which-way information represented by $\mathcal{P}$
and~$\mathcal{Q}$.

Asymmetric interferometers deserve attention because they represent the most
general initial preparation of the Quanton; symmetric interferometers (with $%
\mathcal{P}=0$) are a particular case. Moreover, a number of
proposed which-way experiments are essentially asymmetric: e.g.,
the Einstein recoiling slit in a Young double-slit
interferometer~\cite{Einstein49}, the quantum-optical Ramsey
interferometer outlined in~\cite{Englert92}, and the recent
experiments in~\cite{newHaroche2001}, in which beam splitting is
performed by the quantized cavity-mode of a high-finesse
resonator. In all these cases beam splitting and which-way
detection are provided by the same physical interaction. Thus, the
asymmetry of the beam splitter in such devices---and in the
present treatment---is directly coupled to the ability of the WWD
to get entangled with the atom. Our formalism will prove useful in
understanding the interplay between $\mathcal{P}$, $\mathcal{Q}$,
and~$\mathcal{D}$ stemming from this coupling.

This paper is organized as follows. In section II, we describe the
two-way interferometer setup and review Englert's
formalism~\cite{Englert96}. In order to explain the insight gained
by the introduction of a ``quality'' parameter, we first study the
case of a classical WWD; this is done in Sec.~III, where the WWD
is described as a classical binary communication channel.
Section~IV furnishes the definition of the quality~$\mathcal{Q}$
of a quantum which-way detector and an analysis of its properties.
In Sec.~V we specialize our results to a simple illustrating
example: the Symmetric Quanton-Detecton System.  Finally, we end
in Sec.~VI with a summary and a discussion of the results.

%%%%%%%%%%%%%%%%%%%%%%%%%%%%%%%%%%%%%%%%%%%%%%%%%%%%%%%%%%%%%%%%%%%%%%%%%%%%%
%%%%%%%%%%%%%%%%%%%%%%%%%%%%%%%%%%%%%%%%%%%%%%%%%%%%%%%%%%%%%%%%%%%%%%%%%%%%%%%
\section{\label{sec:2}Duality in Two-Way Interferometers}

\begin{figure}
\includegraphics{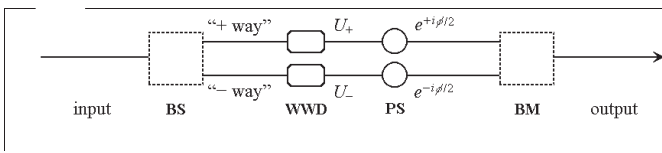}
\caption{\label{fig1}  Schematic two-way interferometer
setup. Beam splitter BS distributes the input states between the 2~ways,
which become entangled with the state of the quantum which-way detector WWD.
Phase shifter PS induces a state-dependent phase shift $\pm \phi /2$. Beam
merger BM recombines the contributions into the final state of the quantum.
Measurements of the output build a fringe pattern versus variation of~$\phi $. }
\end{figure}

Consider the schematic two-way interferometer depicted in Fig.~\ref{fig1}.
Following Englert~\cite{Englert96}, we describe the Quanton degree
of freedom as a spin-$\frac{1}{2}$ system. We prepare the Quanton
in the pure state
\begin{equation}
\rho_Q^{(0)} = \frac{1}{2} \left( 1 + \boldsymbol{s}_Q^{(0)} \cdot \boldsymbol{\sigma}
\right),  \label{2}
\end{equation}
where $\boldsymbol{\sigma} = (\sigma_x,\sigma_y,\sigma_z) $ are
the usual Pauli spin operators. We have chosen an arbitrary
initial pure state with polarization vector
$\boldsymbol{s}_Q^{(0)} = (s_{Qx}^{(0)}, s_{Qy}^{(0)},
s_{Qz}^{(0)}) $, $|\boldsymbol{s}_Q^{(0)}| = 1 $.
The pure state $\rho_Q^{(0)}$ was parameterized in Ref.~\cite{Englert96} as $%
\boldsymbol{s}_Q^{(0)} = (0, -\sin\theta, \cos\theta) $. The
significance of choosing $s_{Qx}^{(0)}\neq 0$ is discussed below.

The detector is prepared in some initial state $\rho_D^{(0)}$ so
that the combined Quanton-plus-detector system is represented~by
\begin{equation}
\rho^{(0)} = \rho_Q^{(0)} \otimes \rho_D^{(0)}.  \label{3}
\end{equation}
As depicted in Fig.~\ref{fig1}, the interferometer consists of a
beam splitter, a phase shifter and WWD for the two~ways, and a
beam merger. The action of the beam splitter on the Quanton is
described by the transformation
\begin{equation}
\rho _{Q}\rightarrow \exp \left( -i\frac{\pi }{4}\sigma _{y}\right)
 \rho_{Q} \; \exp \left( i\frac{\pi }{4}\sigma _{y}\right) ,
\label{4}
\end{equation}
while the beam merger is described by the inverse
transformation\cite {PhaseConv}. In the middle of the
interferometer, the evolution of the detector degree of freedom
after its interaction with the Quanton can be characterized by
\begin{equation}
\rho _{D}\rightarrow U_{\pm }^{\dag } \, \rho _{D} \, U_{\pm }\equiv \rho
_{D}^{(\pm )}\hspace{1cm}\mbox{\rm for the}\;\sigma _{z}=\pm 1\;\mbox{\rm
way},  \label{5}
\end{equation}
where $U_{+}$ and $U_{-}$ are unitary operators acting exclusively
on the detector \cite{nonunitary}. Thus, the unitary operator
governing the evolution of the entire system in the ``split-beam''
section of the interferometer~is
\begin{equation}
\frac{1}{2}(1+\sigma _{z})e^{i\phi /2} \, U_{+}+\frac{1}{2}(1-\sigma
_{z})e^{-i\phi /2} \, U_{-}\,,
\label{6}
\end{equation}
where $e^{\pm i\phi /2}$ are the relative phase factors induced by the phase
shifters. The system's final state is then given by the expression
\begin{eqnarray}
\rho ^{(f)} &=&\mbox{$\frac{1}{4}$}(1-s_{Qx}^{(0)})(1-\sigma_{x}) \;
               U_{+}^{\dag }\rho _{D}^{(0)}U_{+}
\nonumber \\
            &+&\mbox{$\frac{1}{4}$} (1+s_{Qx}^{(0)})(1+\sigma _{x}) \;
                U_{-}^{\dag }\rho _{D}^{(0)}U_{-}
\nonumber \\
            &+&\mbox{$\frac{1}{4}$}(s_{Qz}^{(0)}-is_{Qy}^{(0)})
            (\sigma_{z}+i\sigma_{y})\, U_{+}^{\dag }\rho _{D}^{(0)}U_{-}\;e^{-i\phi }
\nonumber \\
             &+&\mbox{$\frac{1}{4}$}(s_{Qz}^{(0)}+is_{Qy}^{(0)})
             (\sigma_{z}-i\sigma_{y}) \, U_{-}^{\dag }\rho_{D}^{(0)}U_{+}\;e^{i\phi }\,.  \label{7}
\end{eqnarray}
%
%! Note that Eq.~(\ref{7}) reduces to the form given in \cite{Englert96} for
%! $\boldsymbol{s}_Q^{(0)}=(0,0,1)$. $\boldsymbol{s}_Q^{(0)}=(0, -\sin\theta, \cos\theta)$
%! \cite{PhaseConv}.
%
After tracing over the detector degree of freedom, the final state
of the Quanton is described in terms of a Bloch
vector~$\boldsymbol{s}_{Q}^{(f)}$~with
\begin{eqnarray}
&s_{Qx}^{(f)}&= s_{Qx}^{(0)}
\nonumber \\
&s_{Qz}^{(f)}&-is_{Qy}^{(f)} =
(s_{Qz}^{(0)}-is_{Qy}^{(0)})\,\mathcal{C}\,e^{-i\phi }+c.c.,
\label{8}
\end{eqnarray}
where
\begin{equation}
\mathcal{C}\equiv \text{tr}_{D}\left\{ U_{+}^{\dag }\, \rho
_{D}^{(0)} \, U_{-}\right\} \label{9}
\end{equation}
is a complex number---a contrast factor---characterizing \textit{the detector
only}.

To measure interference between the two ways after they are merged, the
observable $\sigma_z$ (or~$\sigma_y$) is measured at the output port of the
interferometer. The probability of finding the value~$\pm1$ for given phase
shift~$\phi$~is
\begin{eqnarray}
p_\pm(\phi) &=& \text{tr}_D \left\{ \frac{1 \pm \sigma_{z}}{2} \rho^{(f)}
\right\} = \frac{1}{2} \left( 1 \pm s_{Qz}^{(f)} \right)
\nonumber \\
&=& \frac{1}{2} \left\{ 1 \pm \text{Re} \left[ ( s_{Qz}^{(0)} - i
s_{Qy}^{(0)} ) \: \mathcal{C} \, e^{-i\phi} \right] \right\}.
\label{10}
\end{eqnarray}
The magnitude of the term in square brackets is the observed
fringe visibility~$\mathcal{V}$. Now suppose the which-way
detector were ``turned off'' ($U_+=U_-$) or even removed
($U_\pm=1$), in which cases $\mathcal{C}=1$. Equation~(\ref{10})
then implies that
\begin{equation}
\mathcal{V}_0 = \left| s_{Qz}^{(0)} - i s_{Qy}^{(0)} \right| = \sqrt{%
(s_{Qy}^{(0)})^2 +(s_{Qz}^{(0)})^2}  \label{11}
\end{equation}
is the \textit{a~priori} visibility of the fringes displayed as
the phase~$\phi$ is varied over repeated runs of the experiment.
When the WWD is ``turned on'' ($U_+ \ne U_-$), the visibility
becomes degraded by a factor~$|\mathcal{C|} $:
\begin{equation}
\mathcal{V} = |\mathcal{C}| \mathcal{V}_0 \,, \hspace{2em} 0\le
|\mathcal{C}|\le 1 . \label{13}
\end{equation}
Therefore, any nontrivial measurement by a WWD invariably results
in a loss of fringe visibility of the Quanton.

An alternative series of measurements on the output could
determine which of the two ways the Quanton has taken on the
average, e.g., by measuring the observable $\sigma _{x}$ (or
$\sigma _{z}$ by removing the beam merger). The probabilities for
taking the $\pm $~way are then given~by
\begin{equation}
w_{\pm }=\text{tr}\left\{ \frac{1\mp \sigma _{x}}{2}\rho ^{(f)}\right\} =%
\frac{1}{2}(1\mp s_{Qx}^{(f)})=\frac{1}{2}(1\mp s_{Qx}^{(0)}),
\label{11.bis}
\end{equation}
respectively. The predictability of the ways is the magnitude of their
difference,
\begin{equation}
\mathcal{P}=|w_{+}-w_{-}|=|s_{Qx}^{(f)}|=|s_{Qx}^{(0)}|\,,
\label{12}
\end{equation}
which is the same whether or not the detector or phase shifter is
operating. The case~$\mathcal{P}=0$ [$s_{Qx}^{(0)}=0$] represents
symmetric
interferometers, where both ways are equally probable, while~$\mathcal{P}=1$ [$%
s_{Qx}^{(0)}=\pm 1$] corresponds to the extreme asymmetric case of
a single-way situation. In terms of which-way information, the predictability $%
\mathcal{P}$ represents the knowledge that the experimenter has
about the ways before measuring $\sigma _{z}$ or $\sigma _{y}$
(i.e., interference effects) on the Quanton. $\mathcal{P}$~is
\textit{a~priori} which-way information stemming from the initial
preparation of the state and the characteristics of the beam
splitter chosen. For instance, the experimenter may opt for a
maximally asymmetric beam splitter (for instance by removing it)
so he or she knows in advance the way to be taken by an atom
prepared in a certain state. In this case it is clear that
$\mathcal{P}=1$ implies $\mathcal{V}=0$, as demanded by duality.
In intermediate cases, the corresponding degradation of the
\textit{a~priori} fringe visibility induced by $\mathcal{P}$ is
implicit in the constraint on the norm of the Bloch
vector~\cite{Englert96}:
\begin{equation}
0\leq
|\boldsymbol{s}_{Q}^{(0)}|^{2}=\mathcal{P}^{2}+\mathcal{V}_{0}^{2}\le
1. \label{14}
\end{equation}
Thus, $\sqrt{1-\mathcal{P}^{2}}$ sets up an absolute upper bound
on the value of \textit{a~priori\/} fringe visibility that can be
measured. Since $\mathcal{V}\le \mathcal{V}_{0}$, the upper bound
on the measured fringe visibility~$\mathcal{V}$ is even less in
the presence of a~WWD.

%% Now turn on the which-way detector.  Inserting  (\ref{11}) into (\ref{10}),
%% we see that the {\em a~priori} fringe visibility is degraded by a factor
%% $|\mathcal{C}|$, namely,  $\ldots\,$.
%% %
%% %! \begin{equation}
%% %! %
%% %! \mathcal{V}= |\mathcal{C}| \mathcal{V}_0 , \hspace{1 cm} 0\le |\mathcal{C}|\le 1  .
%% %! \label{13}
%% %! %
%% %! \end{equation}
%% %
%% %
%% Thus, the acquisition of which-way information by a quantum detector
%% described by Eq.\ (\ref{5}) results in a degradation of the final
%% fringe visibility of the Quanton, as required by duality.

Summing up, we can say that there are two different sources of
which-way information resulting in degradation of the fringe
visibility. One is the \textit{a~priori} which-way knowledge
characterized by~$\mathcal{P}$. The second source of which-way
information arises from the \textit{quantum properties of the
detector}, through the correlations established after its
entanglement with the Quanton at the central stage of the
interferometer. A quantitative measure of the total
distinguishability of the ways as determined by the detector is
afforded by the quantity \cite{Englert96}
\begin{equation}
\mathcal{D} \equiv \text{tr}_D \left\{ \left| w_+ \rho_D^{(+)}-
w_- \rho_D^{(-)} \right| \right\} \ge \mathcal{P},  \label{15}
\end{equation}
where $\rho_D^{(+)}$, $\rho_D^{(-)}$ are the detector's two final states~(%
\ref{5}) corresponding to each way. These compose the final detector state
according~to
\begin{equation}
\rho_D^{(f)} = \text{tr}_Q \, \rho^{(f)}  = w_+ \rho_D^{(+)} +
w_- \rho_D^{(-)} .
\label{16}
\end{equation}
As shown in Ref.~\cite{Englert96}, (\ref{13}) and (\ref{15})
always satisfy the inequality (\ref{1}) quantifying the extent to
which partial which-way information and partial fringe visibility
are compatible. This statement of duality generalizes the
inequality~(\ref{14}) to include the detector's role in storing
knowledge of the Quanton's initial state.

%%%%%%%%%%%%%%%%%%%%%%%%%%%%%%%%%%%%%%%%%%%%%%%%%%%%%%%%%%%%%%%%%%%%%%
%%%%%%%%%%%%%%%%%%%%%%%%%%%%%%%%%%%%%%%%%%%%%%%%%%%%%%%%%%%%%%%%%%%%%%%%%
%
\section{\label{sec:classical}The ``Quality'' of a Classical Binary Channel}

\begin{figure}
\includegraphics{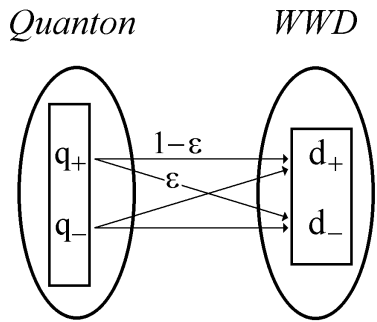}
\caption{\label{fig2}  Classical Binary Channel. Correlations
between Quanton and WWD are regarded as a classic communication
problem between two binary systems, sender {\it Quanton} and
receiver {\it WWD} with state values ($q_{\text{+}},q_{-}$) and
($d_{\text{+}},d_{-}$) respectively. The communication is
established trough a noisy channel with error probability
$\epsilon$. }
\end{figure}

In order to explain the meaning of a quality parameter for a
quantum WWD, we study first the case of a classical WWD, i.e., a
WWD which can only establish classical correlations with the
Quanton. The information gained by the WWD can be described as a
problem of classical communication between a sender
\textit{Quanton} and a receiver \textit{WWD} through a noisy
channel~\cite {Gallager}. A general diagram of the channel is
shown in Fig.~\ref{fig2}.
Here $q_\pm$ are the two possible states of the Quanton. Classical
correlations are established with a binary
 \textit{WWD} with two readouts, $d_+$ and~$d_-$, through a noisy channel characterized by a
probability of error $\epsilon$. The joint probabilities defining the communication system are~\cite{epsilon} %
\begin{equation}
\begin{array}{rcl}
P_{QD}(q_+,d_+) & = & w_+ (1-\epsilon) \,, \\
P_{QD}(q_+,d_-) & = & w_+ \epsilon \,, \\
P_{QD}(q_-,d_+) & = & w_- \epsilon \,, \\
P_{QD}(q_-,d_-) & = & w_- (1-\epsilon) \,.
\end{array}
\label{2.1}
\end{equation}
Following the notation of the previous section, we have written the
probabilities of each alternative in the sender as
\begin{equation}
P_{Q}(q_{\pm}) = w_{\pm} \,.  \label{2.2}
\end{equation}
The predictability of the ways, $\mathcal{P} = |w_+-w_-| $
[Eq.~(\ref{12})], gives the dispersion of the probability
distribution. It may also be written
\begin{equation}
\mathcal{P} = \sqrt{ 2\left< P \right> -1 } \:,  \label{2.3}
\end{equation}
where $\left< P \right> = [P_Q(q_+)]^2 + [P_Q(q_-)]^2 $ is the mean value of
the probability itself ($\frac12 \le \left<P\right> \le 1$).

We can therefore understand $\mathcal{P}$ as a measure of the
information content of the probability distribution, and can
further connect this information with a betting strategy of
Englert~\cite{Englert96}. Consider a measurement of the Quanton
subsystem by the \textit{WWD}\@. Before an outcome is registered,
we can bet on which alternative is going to occur. The
likelihood~$\mathcal{L}$ of our guessing right is connected to the
distinguishability of the results via
\begin{equation}
\mathcal{L} \equiv \frac{1}{2} \left( 1+\mathcal{D} \right) .
\label{2.3bis}
\end{equation}
This means that when $\mathcal{D}=1$ the ways can be totally
distinguished and we can win the bet 100\% of the time, whereas
$\mathcal{D}=0$ implies that
there is no knowledge about the ways. Note that in the latter case (with $%
\mathcal{L}=\frac12$) we can still win 50\% of the time, as in a
coin toss, because even here we have some information about the
system: we know in advance that the Quanton is a two-level system
so we are right about the ways as often as~not.

If we make a measurement of the ``sender'' Quanton directly, as
characterized by Eq.~(\ref{2.1}), the best bet is to commit to the
alternative occurring with the greatest probability. The
likelihood of guessing correctly follows from (\ref{2.2}):
\begin{equation}
\mathcal{L}=\text{Max}\{w_{+},w_{-}\}=\frac{1}{2}(1+\mathcal{P}).
\label{2.4}
\end{equation}
Equation~(\ref{2.4}) reveals $\mathcal{P}$ as the classical
\textit{a~priori} distinguishability of the ways: $\mathcal{P}=1$
indicates full \textit{a~priori} knowledge of the ways, while the
maximum uncertainty occurs for $\mathcal{P}=0$. The alternatives
are equally probable when $\mathcal{P}=0$ ($w_{+}=w_{-}$) and the
ways cannot be distinguished at~all.

Now we quantify the classical which-way information that can be acquired
when we turn the ``receiver'' \textit{WWD} on. The information about the
occurrence of $q_{\pm }$ acquired by reading an outcome $d_{\pm }$ is given
by the conditional probabilities
\begin{equation}
\begin{array}{rcccl}
P_{D/Q}(d_{+}/q_{+}) & = & P_{D/Q}(d_{-}/q_{-}) & = & (1-\epsilon )\,,%
\nonumber \\
P_{D/Q}(d_{+}/q_{-}) & = & P_{D/Q}(d_{-}/q_{+}) & = & \epsilon \,.
\end{array}
\label{2.5}
\end{equation}
The distance between the two conditional probabilities leading to
the same outcome in the \textit{WWD} tells us how noisy the
channel~is. This motivates the following definition of the quality
$\mathcal{Q}$ of the channel:
\begin{equation}
\mathcal{Q} \equiv \left| P_{D/Q}(d_{\pm }/q_{+})-P_{D/Q}(d_{\pm
}/q_{-})\right| =1-2\epsilon \label{2.6}
\end{equation}
The case $\epsilon =0$ characterizes a noiseless channel, for
which the subsystems \textit{Quanton} and \textit{WWD} are
perfectly correlated and the quality is maximal ($\mathcal{Q}=1$).
In this case, the alternatives of the Quanton are totally
distinguishable by reading the outcomes of the \textit{WWD} . In
the opposite extreme, $\epsilon =\frac{1}{2}$ represents a
maximally noisy channel of the poorest quality ($\mathcal{Q}=0$)
and the outcomes in the sender and receiver are totally
uncorrelated. A reading of the \textit{WWD} cannot increase our
knowledge of the alternatives of the Quanton at all. From this it
follows that $\mathcal{Q}$ must contribute to the
distinguishability of the ways, once the \textit{WWD} has been
turned~on.

In order to quantify the contribution of~$\mathcal{Q}$ to the
distinguishability, let us measure some property~$f$ of the
\textit{WWD}. The mean value of that property~is
\begin{equation}
\langle f\rangle =f(d_{+})\,P_{D}(d_{+})+f(d_{-})\,P_{D}(d_{-}).
\label{2.7}
\end{equation}
The probabilities $P_{D}(d_{\pm})$ for each outcome in {\it WWD}
can be calculated by summing the two contributions leading to each
possible outcome $d_{\pm }$. From Eqs.~(\ref{2.1}) we obtain
\begin{eqnarray}
P_{D}(d_{+}) &=&w_{+}(1-\epsilon )+w_{-}\epsilon ,  \nonumber  \label{2.8} \\
P_{D}(d_{-}) &=&w_{+}\epsilon +w_{-}(1-\epsilon ).
\end{eqnarray}
After an outcome $d_{\pm }$ in {\it WWD} is obtained, we can bet
on the Quanton's alternative that contributes most to the
probability $P_{D}(d_{\pm })$. As in (\ref{2.4}), after many
repetitions this betting procedure yields the likelihood for
guessing right:
\begin{equation}
\mathcal{L}=\text{Max}\{w_{+}(1-\epsilon ),w_{-}\epsilon \}+\text{Max}%
\{w_{+}\epsilon ,w_{-}(1-\epsilon )\}.  \label{2.9}
\end{equation}
Comparing Eqs. (\ref{2.9}) and (\ref{2.3bis}), and using the identity
\begin{equation}
\text{Max}\{x,y\}=\frac{1}{2}(x+y)+\frac{1}{2}|x-y|,\hspace{1cm} \forall x,y ,
\label{2.10}
\end{equation}
we obtain the result
\begin{equation}
\mathcal{D}=\text{Max}\{\mathcal{P},\mathcal{Q}\}. \label{2.11}
\end{equation}
Summarizing, we can say that there are two sources of which-way
information in the communication system. One is the a-priori
distinguishability $\mathcal{P} $ inherent in the preparation of
the Quanton system previous to the interaction with the {\it WWD}.
The other is the distinguishability $\mathcal{Q} $ provided by the
{\it WWD}. As stated in Eq.~(\ref{2.11}), the total
distinguishability $\mathcal{D}$ is given by the greater of the
two. The
situation turns out to be far more complicated when we quantize the {\it WWD}%
, as will be shown in the following section.

%%%%%%%%%%%%%%%%%%%%%%%%%%%%%%%%%%%%%%%%%%%%%%%%%%%%%%%%%%%%%%%%%%%%%%%%%%%%%%
%%%%%%%%%%%%%%%%%%%%%%%%%%%%%%%%%%%%%%%%%%%%%%%%%%%%%%%%%%%%%%%%%%%%%%%%%%%%%%%%%

\section{\label{sec:quantum}The ``Quality'' of a Quantum WWD}
Consider that the Quanton interferometer is equipped with a
quantum WWD, in the fashion described in Section I. In order to
isolate the contribution to the distinguishability arising solely
from the quantum properties of the detector, we define the
``quality'' of the quantum WWD to be
\begin{equation}
\mathcal{Q\equiv }\frac{1}{2}\text{tr}_{D}\{|\rho _{D}^{(+)}-\rho
_{D}^{(-)}|\}. \label{18}
\end{equation}
Thus, $\mathcal{Q}$ coincides with~$\mathcal{D}$ in the case of
symmetric interferometers ($\mathcal{P}=0,w_{\pm }=\frac{1}{2}$).
On the other hand, in contrast to (\ref{15}), $\mathcal{Q}$ does
not involve the \textit{a~priori} probabilities of the ways
represented by $w_{\pm }$, so both quantities may differ
substantially in the case of asymmetric interferometers. Note
that, as in the previous section, $\mathcal{Q}$ is the distance
between two conditional probabilities, $\rho _{D}^{(+)}$ and $\rho
_{D}^{(-)}$, in the trace-class norm, and thus it is a
quantitative measure of the detector's intrinsic ability to
distinguish between the ways. The detector cannot distinguish
between the ways at all if $\mathcal{Q}=0$ and, conversely, full
which-way information can be extracted by the detector when
$\mathcal{Q}=1$. The states $\rho _{D}^{(\pm )}$ can be prepared
by means of measuring the actual way taken by the atom. Thus, the
value of~$\mathcal{Q}$ can be experimentally checked along the
lines described in \cite{Englert96} for measuring $\mathcal{D}$.

To establish a relation between (\ref{15}) and (\ref{18}), we first consider
the detector to be prepared in a pure state, so that the equality holds in (%
\ref{1}). Then (\ref{18}) is a distance between projectors, which
can be easily calculated to yield
\begin{equation}
\mathcal{Q}^2 +|\mathcal{C}|^2 = 1.  \label{19}
\end{equation}
Comparing (\ref{19}) with (\ref{1}) and using (\ref{13}) we obtain
\begin{equation}
\mathcal{Q}^2=\frac{\mathcal{D}^2-\mathcal{P}^2}{1-\mathcal{P}^2}
\le \mathcal{D}^2\le 1. \label{20}
\end{equation}
Thus, for pure state preparation, the different distinguishability
measures satisfy $\mathcal{D}\geq \mathcal{Q}$, $\mathcal{D}\geq
\mathcal{P}$.

Consider next the general case in which we allow the detector to be in a
mixed state. In analogy to (\ref{1}), the three quantities $\mathcal{Q}$, $%
\mathcal{P}$ and $\mathcal{V}$ should be related by an inequality
that is an expression of duality. We can obtain this inequality in
a straightforward fashion by noticing that the derivation of
(\ref{1}) presented in \cite {Englert96} also applies to our case
under the replacements
\begin{equation}
\mathcal{D} \rightarrow \mathcal{Q}, \qquad \mathcal{V}
\rightarrow |\mathcal{C}|, \label{21}
\end{equation}
which transform (\ref{1}) into
\begin{equation}
\mathcal{Q}^2 +|\mathcal{C}|^2 \le 1.  \label{22}
\end{equation}
Then inserting (\ref{13}) and (\ref{14}) into (\ref{22}) gives
\begin{equation}
\left( 1-\mathcal{P}^2 \right)\mathcal{Q}^2 +\mathcal{P}^2
+\mathcal{V}^2 \le 1 . \label{23}
\end{equation}
This equation, which is a quantitative statement about duality, constitutes
the central result of this paper.

\begin{figure}
\includegraphics{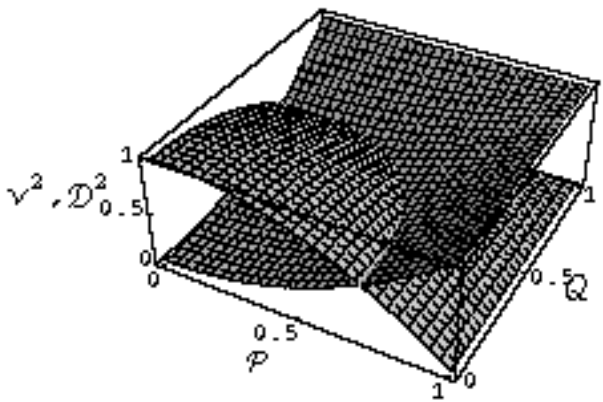}
\caption{\label{figpico} Duality in the pure state preparation
case. We plot  $\mathcal{V}^2$ and $\mathcal{D}^2$, satisfying
$\mathcal{D}^2 +\mathcal{V}^2=1$, as a function of $\mathcal{P}$
and $\mathcal{Q}$. Note that both quantities are invariant under
the permutation $\mathcal{Q} \leftrightarrow \mathcal{P}$. }
\end{figure}

Extreme situations characterized by perfect fringe visibility or perfect
which-way information are mutually exclusive. Thus, duality demands

\begin{subequations}
\label{boncho}
\begin{eqnarray}
\mathcal{V}=1 & \quad\Rightarrow\quad &
\mathcal{D}=\mathcal{P}=\mathcal{Q}=0 ,
\label{1.1bis} \\
\mathcal{P}=1 & \quad\Rightarrow\quad & \mathcal{V}=0,  \label{1.2bis} \\
\mathcal{D}=1 & \quad\Rightarrow\quad & \mathcal{V}=0,  \label{1.3bis} \\
\mathcal{Q}=1 & \quad\Rightarrow\quad & \mathcal{V}=0 .
\label{1.4bis}
\end{eqnarray}
\end{subequations}
It is easy to check that the extreme situations described in (\ref{1.1bis}%
)--(\ref{1.4bis}) devolve from Eq.\ (\ref{23}). In particular, the
condition $\mathcal{P}=1$ exhausts the amount of which-way
information that can be available about the Quanton so
(\ref{1.2bis}) has to be satisfied whatever the value
of~$\mathcal{Q}$ is. Conversely, (\ref{1.4bis}) has to be
satisfied for arbitrary values of~$\mathcal{P}$. This feature is
contained in the structure of the left hand side of Eq.\
(\ref{23}), which does not involve the value of~$\mathcal{Q}$ in
the case that $\mathcal{P}$ becomes maximum, and vice-versa. Note,
moreover, that Eq.\ (\ref{23}) is invariant under the permutation
$\mathcal{Q} \leftrightarrow \mathcal{P}$, which is clear from its
alternative form
\begin{equation}
\mathcal{V}^2\le (1-\mathcal{P}^2) (1-\mathcal{Q}^2).
\label{3.6bis}
\end{equation}
Consequently, as far as the degradation of the Quanton's fringe
visibility is concerned, both sources of which-way information
stand on an equal footing in situations where (\ref{3.6bis}) is
satisfied with the equal sign. This symmetry can be appreciated in
Fig. \ref{figpico} where $\mathcal{D}^2$ and $\mathcal{V}^2$ are
plotted as a function of $\mathcal{P}$ and $\mathcal{Q}$ in the
pure state preparation case.

In the case where we also allow the Quanton to be initially in a
mixed state, we get the totally general inequality
\begin{equation}
\left( |\boldsymbol{s}_{Q}^{(0)}|^{2}-\mathcal{P}^{2}\right) \mathcal{Q}^{2}+|\boldsymbol{s}%
_{Q}^{(f)}|^{2}\le |\boldsymbol{s}_{Q}^{(0)}|^{2},  \label{23.bis}
\end{equation}
where $|\boldsymbol{s}_{Q}^{(f)}|$ is the norm of the final
Quanton's Bloch vector,
$\boldsymbol{s}_{Q}^{(f)}=\text{tr}_{QD}\{\rho
^{(f)}\boldsymbol{\sigma}_{Q}\}$. Which-way information is stored
in the detector due to its entanglement with the Quanton. The
expression (\ref{23.bis}) offers us a link relating~$\mathcal{Q}$
to the degree of purity left in the Quanton on account of this
entanglement. The magnitude of the entanglement can be measured by
the norm of the Quanton's Bloch vector, which satisfies
\begin{eqnarray}
|\boldsymbol{s}_{Q}|^{2} &=&\mathcal{P}^{2}+\mathcal{V}^{2}  \nonumber \\
&=&1+2\;\text{tr}\left\{ \rho _{Q}^{2}-\rho _{Q}\right\} \le 1,
\label{24}
\end{eqnarray}
where $\rho _{Q}=\text{tr}_{D}\, \rho$. The first equality in
(\ref{24}) follows from Eqs.~(\ref{8}), (\ref{11}) and (\ref{12}),
the second equality follows trivially from (\ref{2}) and the
properties of the Pauli matrices. The decrease in the norm of the
Bloch vector measures the degree of deviation of the Quanton from
a pure state. In fact, the bounds in (\ref{13}) guarantee that the
entropy-like quantity \cite{Gallis96}
\begin{equation}
\mathcal{G}_{Q}\equiv 1-\text{tr}\{\rho _{Q}^{2}\}=\frac{1}{2}(1-|\boldsymbol{s}%
_{Q}|^{2})  \label{25}
\end{equation}
always increases when the Quanton's fringes degrade as a result of
its
entanglement with the detector. More explicitly, inserting (\ref{13}) and (%
\ref{24}) into (\ref{25}) we find
\begin{equation}
\Delta \mathcal{G}_{Q}\equiv \mathcal{G}_{Q} -
\mathcal{G}_{Q}^{(0)} = \frac{1}{2}\left(
\mathcal{V}_{0}^{2}-\mathcal{V}^{2}\right) . \label{26}
\end{equation}
The inequality given in (\ref{23.bis}) can now be recast in terms
of the ``linear entropy'' (\ref{25}) as
\begin{equation}
\mathcal{Q}^{2}\le \frac{2 \Delta \mathcal{G}_{Q}
}{\mathcal{V}_{0}^{2}}\le 1. \label{27}
\end{equation}
The implications to be drawn from Eq.\ (\ref{27}) are
straightforward. First, we see that  $\mathcal{Q}=0$ is obtained
for $\mathcal{G}_{Q}^{(0)}=\mathcal{G}_{Q}$, i.e., degradation of
the purity of the state of the Quanton is a necessary condition
for the extraction of quantum WWI about the Quanton alternatives
($\mathcal{Q}\neq 0$). Conversely, according to Eq. (\ref{27}),
maximal $\mathcal{Q}=1$ can only occur when $\mathcal{G}_{Q}$ is
maximized. In this case $\mathcal{V}=0$, so that
$|\boldsymbol{s}_{Q}|$ for the Quanton in Eq.~(\ref{24}) has been
maximally degraded from $|\boldsymbol{s}_{Q}^{(0)}|$. Moreover, we
can say that $\mathcal{Q}=1$ indicates optimal which-way detection
at the expense of the total destruction of the fringe pattern
($\mathcal{V}=\mathcal{C}=0$). In intermediate situations, the
amount of which-way information that can be extracted by the
detector is bounded by the degree of purity lost by the Quanton.

%
%%%%%%%%%%%%%%%%%%%%%%%%%%%%%%%%%%%%%%%%%%%%%%%%%%%%%%%%%%%%%%%%%%%%%%%%%
%

\section{\label{sec:SQDS}The Symmetric Quanton-Detecton System}
\begin{figure}
\includegraphics{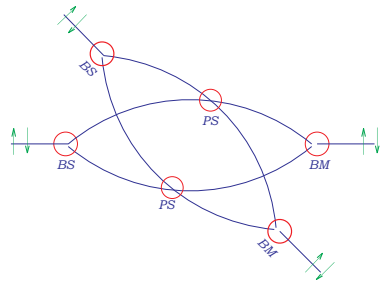}
\caption{\label{figojo}  Schematic setup for the SQDS, showing the
intrinsic symmetry between Quanton and Detecton. }
\end{figure}

We consider in this section a particular model for the WWD. A
2-state detector, or Detecton, is the simplest possible quantum
device that can probe WWI about the Quanton. The Detecton, as the
Quanton itself, is an 2-ways interferometer, likewise describable
by a predictability  $\mathcal{P}_D$, and a fringe visibility
$\mathcal{V}_D$. Its initial state can also be described by a
Bloch vector $\boldsymbol{s}_{D}^{(0)}$, which can be subjected to
the transformations described in Section \ref{sec:2}, i.e., BS,
phase shift $\phi_D$ and BM, as the original Quanton. Both
interferometers are assumed to interact at their central stages,
where they become entangled according to Eq. (\ref{6}). Since both
interferometers can play the role of System or WWD of each other,
two parameters $\mathcal{Q}_D$, $\mathcal{Q}_Q$ have to be
introduced to measure Detecton큦 and Quanton큦 "qualities" as
which way detectors, respectively. Actually, we have designed the
device so as to the system becomes entirely symmetric between the
labels Q and D, as can be seen in Fig. \ref{figojo}. Hence, we
will call it the Symmetric Quanton-Detecton System (SQDS)
\cite{JML00}.

The SQDS is a good illustrating example. First, it allows for a
considerable simplification of the formalism. Second, it provides
further insight into the reciprocal effects of Duality between
Quanton and Detecton. Third, it is an interesting system by
itself. Actually, a pair of entangled two-level system (or qubits)
form the fundamental brick in the construction of quantum logic
gates. The SQDS is a quantum logic gate where each qubit, system
and control, play the role of Quanton or WWD of the other.

For the purposes of this Section, The SQDS  can be regarded as two
interferometers trying to acquire WWI about each other. In fact,
the SQDS is in essence a pair of two-ways interferometers coupled
at their central stage by a dispersive interaction.
We take the
Detecton phase shifter to depend on the ways in the form
\begin{eqnarray}
U_{PS}^{\pm}  &=& \exp \left[  \frac{i}{2} \left( \phi_D \pm \Phi \right)  \sigma_{Dz}\right],
\nonumber \\
&=& \frac{1}{2}(1+\sigma _{Dz})e^{i\phi_D /2} e^{\pm i\Phi /2}\,
\nonumber \\
&+& \frac{1}{2}
(1-\sigma_{Dz}) e^{-i\phi_D /2}e^{\mp i\Phi /2} \, .
\label{SQDS2}
\end{eqnarray}
The phase shifts $\left( \phi_Q ,\phi_D ,\Phi \right)$ represent
three arbitrary parameters externally controlling the SQDS.
However, due to the simplicity of the system, entanglement is
controlled in the SQDS by a single parameter, the entangling phase
$\Phi$. For $\Phi=0$ $(\text{mod}\, \pi)$ both systems are
disentangled, $\rho=\rho_Q \otimes \rho_D$. In fact, Eq.
(\ref{SQDS2}) can be regarded as a quantization of the phase
shifter, since the direction of rotation of the angle $\Phi$
performed by one subsystem depends on the way chosen by the other
subsystem. This dispersive interaction provides the conditional
dynamics that lies at the heart of any quantum logic gate.

In order to calculate $\mathcal{Q}_D$, we first take the initial
state of the Detecton as
\begin{equation}
\rho_D^{(0)} = \frac{1}{2} \left( 1 + \boldsymbol{s}_D^{(0)} \cdot
\boldsymbol{\sigma}_D \right), \hspace{0.5cm}
\boldsymbol{s}_D^{(0)} = (s_{Dx}^{(0)},0,s_{Dz}^{(0)}) ,
\label{SQDS3}
\end{equation}
where $|s_{Dx}^{(0)}|=\mathcal{P}_D$,
$\mathcal{V}_D^o=|s_{Dz}^{(0)}|$ and $0\leq |\boldsymbol{s}_D|\leq
1$. For the sake of simplicity, we have taken $s_{Dy}^o$ to
vanish. As a consequence of the $Q\leftrightarrow D$ symmetry, the
BS and BM for the Detecton have the same forma that their Quanton
counterparts given in (\ref{4}). Thus, the action of the BS on
(\ref{SQDS3}) generates the state
\begin{equation}
\rho_D^{BS} = \frac{1}{2} \left( 1 + s_{Dz}^{(0)} \sigma_{Dx} -
s_{Dx}^{(0)} \sigma_{Dz} \right). \label{SQDS4}
\end{equation}
Now, inserting Eq. (\ref{SQDS2}) and (\ref{SQDS4}) into (\ref{5}), the Detecton evolution can be calculated as
\begin{equation}
\rho_D^{\pm} = \frac{1}{2} \left( 1 + \boldsymbol{s}_D^{\pm}\cdot  \boldsymbol{\sigma}_D \right),
\label{SQDS5}
\end{equation}
with
\begin{equation}
\boldsymbol{s}_{D}^{\pm} = \left( s_{Dz}^{(0)} \cos
\varphi_{\pm},s_{Dz}^{(0)} \sin \varphi_{\pm},-s_{Dx}^{(0)}
\right), \label{SQDS6}
\end{equation}
where we have defined the auxiliary variable $\varphi_{\pm}=
\phi_D \pm \Phi$. With the help of (\ref{SQDS5}), and taking into
account the unitary character of (\ref{SQDS2}), we write the
Quality in (\ref{18}) as
\begin{eqnarray}
\mathcal{Q} &=& \frac{1}{2}\text{tr}_{D}\{| \left(
\boldsymbol{s}_{D}^{+}- \boldsymbol{s}_{D}^{-} \right)
\vec{\sigma}_D|\}
\nonumber\\
&=& \frac{1}{2}| \boldsymbol{s}_{D}^{+}- \boldsymbol{s}_{D}^{-} |.
\label{SQDS7}
\end{eqnarray}

Inserting (\ref{SQDS6}) into (\ref{SQDS7}) we arrive at the result
\begin{equation}
\mathcal{Q}_D = \mathcal{V}_D^o |\sin \Phi|. \label{SQDS8}
\end{equation}
We see that in order to get maximum quality of the WWD two
conditions are required. First, we need maximum entanglement
$(\Phi=\frac{\pi}{2}\, \text{mod}(\pi))$. Second, maximum initial
visibility of the Detecton interferometer $(\mathcal{V}_D^o=1)$.
As will be shown later, the Detecton acquires WWI about the
Quanton by degrading its visibility. Thus, maximum storage of WWI
requires maximum initial visibility $\mathcal{V}_D^o$. Thus,
Detectons in a mixed state cannot act as perfect-quality which-way
detectors. On the other hand, as $\mathcal{P}_D^2
+\mathcal{V}_D^{o2} \leq 1$, any amount of predictability in the
Detecton limits its quality. Another interesting feature in
(\ref{SQDS8}) is that $\mathcal{Q}_D$ does not depend on the
predictability nor on the initial state of the Quanton. Full WWI
can be stored in the state of the Detecton for any qubit state
sent through the Quanton interferometer. This feature makes the
SQDS an interesting candidate for a quantum logic gate as will be
shown in a forthcoming publication.

In order to compute Englert큦 distinguishability $\mathcal{D}_Q$
\cite{label}, we define first the operator
\begin{equation}
\underline{\Delta} \equiv  \omega_+ \rho_D^+ -\omega_- \rho_D^- ,
\label{SQDS9}
\end{equation}
so $\mathcal{D}_Q=\text{tr} |\underline{\Delta}|$. In contrast to
the $\mathcal{P}_Q=0$ case, $\underline{\Delta}$ is not traceless
and the diagonalization is more involved, but nevertheless
straightforward. We obtain
\begin{equation}
\mathcal{D}_Q = \text{Max} \left\{  \mathcal{P}_Q, \mathcal{R}_Q
\right\} , \label{SQDS10}
\end{equation}
where
\begin{eqnarray}
\mathcal{R}_Q &=& |\omega_+ \boldsymbol{s}_D^+  -\omega_-
\boldsymbol{s}_D^- |
\nonumber \\
&=&\sqrt{\mathcal{P}_Q^2 \mathcal{P}_D^2 + \mathcal{V}_D^{o2}
\left( \sin^2 \Phi +\mathcal{P}_Q^2 \cos^2 \Phi \right)}  .
\label{SQDS11}
\end{eqnarray}
Eq. (\ref{SQDS10}) is very illustrative. First, note that for
$\mathcal{P}_Q=0$, $\mathcal{D}_Q=\mathcal{R}_Q=\mathcal{Q}_D$. In
this case, there is only one source for Quanton큦 WWI, so the
distinguishability is directly given by the Quality of the WWD. As
we show in this paper, the introduction of a non-vanishing
predictability $\mathcal{P}_Q$ turns out the analysis of
distinguishability much more involved, as can be seen by comparing
Eq. (\ref{SQDS8}) to (\ref{SQDS10}). Second, it can be seen that
Eq. (\ref{SQDS11}) assures the inequality $\mathcal{D}_Q \geq
\mathcal{P}_Q$ given in (\ref{15}). Third, for a decoupled systems
(i.e., $\Phi=0$), Eq. (\ref{SQDS11}) reduces to
$\mathcal{R}_Q=\mathcal{P}_Q |\boldsymbol{s}_D^o| \leq
\mathcal{P}_Q$. Therefore we recover in this case the result
$\mathcal{D}_Q=\mathcal{P}_Q$: there is no WWD available to
increase distinguishability above the \textit{a-priori} WWI.
Consider now the case of maximal coupling ($\Phi=\frac{\pi}{2}$,
mod($\pi$)). In this case $\mathcal{R}_Q=\sqrt{\mathcal{P}_Q^2
\mathcal{P}_D^2 + \mathcal{V}_D^{o2}}$. Thus, $\mathcal{V}_D^o=1$
is required in order to increase $\mathcal{D}_Q$ from
$\mathcal{P}_Q$ to $1$. Thus, as commented before after
calculation of $\mathcal{Q}_D$, the existence of a non-vanishing
$\mathcal{P}_D$ acts as a limiting factor for WWI storage
capability. Fourth, notice that Eq. (\ref{SQDS11}) can be
rewritten as
\begin{equation}
\mathcal{R}_Q^2 = \mathcal{P}_Q^2 |\boldsymbol{s}_D^o|^2 +
\mathcal{Q}_D^2
 \left( 1-\mathcal{P}_Q^2  \right) .
\label{SQDS11rewritten}
\end{equation}
Therefore, for $|\boldsymbol{s}_D^o|=1$ we obtain  $\mathcal{R}_Q
\geq \mathcal{P}_Q$ and $\mathcal{D}_Q=\mathcal{R}_Q$ in this
case. Thus, $\mathcal{R}_Q$ gives the total distinguishability for
initial pure state preparation of the Detecton. Also, since
$\mathcal{Q}_D \leq |\boldsymbol{s}_D^o|$, Eq.
(\ref{SQDS11rewritten}) implies $\mathcal{R}_Q \leq
|\boldsymbol{s}_D^o|$, i.e., $\mathcal{R}_Q$ is bounded by the
initial purity of the Detecton. On the other hand, it is easy to
show from Eq. (\ref{SQDS11rewritten}) that $\mathcal{R}_Q \geq
\mathcal{Q}_D$. Thus, in the SQDS, the following inequalities are
satisfied with generality
\begin{equation}
\mathcal{D}_Q \geq \mathcal{R}_Q \geq \mathcal{Q}_D, \hspace{1cm}
\mathcal{D}_Q \geq \mathcal{P}_Q , \label{SQDS11bis}
\end{equation}
setting a hierarchy for the different distinguishability measures.
This leads trivially to the following a hierarchy of duality
inequalities
\begin{subequations}
\label{boncho2}
\begin{eqnarray}
\mathcal{Q}_D^2 +\mathcal{V}_Q^2 &\leq& \mathcal{R}_Q^2
+\mathcal{V}_Q^2 \leq \mathcal{D}_Q^2 +\mathcal{V}_Q^2 \leq 1 ,
\label{boncho2.1} \\
\mathcal{P}_Q^2 +\mathcal{V}_Q^2 &\leq& 1 , \label{boncho2.2}
\end{eqnarray}
\end{subequations}
where we have used Eq. (\ref{1}) in (\ref{boncho2.1}) and Eqs.
(\ref{13}) and (\ref{14}) in (\ref{boncho2.2}).

\begin{figure}
\includegraphics{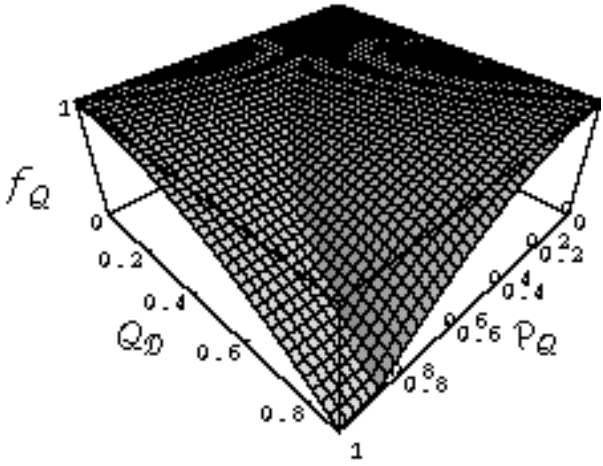}
\caption{\label{figf}  $f_Q$ [Eq. (\ref{SQDSf})] for
$|\boldsymbol{s}_Q^o|=0.882$, as a function of $\mathcal{P}_Q$ and
$\mathcal{Q}_D$. }
\end{figure}

Now, we are in conditions to show an interesting feature. In the
SQDS, the inequality (\ref{3.6bis}) is more stringent than the
Englert큦 inequality (\ref{1}). In order to show this, we define
the auxiliary quantity
\begin{equation}
f_Q \equiv
\frac{(1-\mathcal{P}_Q^2)(1-\mathcal{Q}_D^2)}{(1-\mathcal{D}_Q^2)}
. \label{SQDSf}
\end{equation}
For $|\boldsymbol{s}_D^o|=1$, $f_Q=1$ and both inequalities are
equivalent. Also, $f_Q=1$ follows trivially from
$\mathcal{P}_Q=0$, or $\mathcal{Q}_D=0$, or $\mathcal{Q}_D=1$. In
the general case, we show in the appendix that $f_Q \leq 1$,
showing the validity of our stringency statement. A typical
behavior of $f_Q$ is given in Fig. \ref{figf}, where Eq.
(\ref{SQDSf}) is plotted for $|\boldsymbol{s}_Q^o|=0.882$ versus
different values of $\mathcal{P}_Q$ and $\mathcal{Q}_D\leq
|\boldsymbol{s}_Q^o|$. As can be seen in this plot, there are two
behaviors, corresponding to two different regions of the parameter
space where $\mathcal{R}_Q
> \mathcal{P}_Q$ (right) or $\mathcal{R}_Q < \mathcal{P}_Q$ (left).

In order to see how all these parameters are related to fringe
degradation, we compute now $\mathcal{V}_Q$ as given in
(\ref{13}). Inserting (\ref{SQDS2}) into (\ref{9}), we obtain
\begin{equation}
\mathcal{C} = \cos \Phi +is_{Dx}^o \sin \Phi . \label{SQDS12}
\end{equation}
Therefore
\begin{equation}
\mathcal{V}_Q = \mathcal{V}_Q^o \,\sqrt{\cos^2 \Phi
+\mathcal{P}_D^2 \sin^2\Phi } . \label{SQDS13}
\end{equation}

Now, we are in condition to illustrate several inequalities
involving duality that appeared in the previous section. For
instance, for pure state preparation we compute Eq. (\ref{19}) as
\begin{eqnarray}
\mathcal{Q}_D^2 +|\mathcal{C}|^2 &=&
\left(1-\mathcal{P}_D^2\right) \sin^2\Phi +\cos^2\Phi+
\mathcal{P}_D^2 \sin^2\Phi
\nonumber \\
&=& \cos^2\Phi +\sin^2\Phi =1 .
\label{SQDS14}
\end{eqnarray}
As can be seen from the above equation, the introduction of
$\mathcal{P}_D$ decreases the Quality of the WWD and increases the
contrast factor by the same amount, so the sum remains constant.
For the general case we compute
\begin{equation}
\mathcal{Q}_D^2 +|\mathcal{C}|^2 = |\boldsymbol{s}_D^o|^2
\sin^2\Phi + \cos^2\Phi \leq 1 , \label{SQDS15}
\end{equation}
so we see that mixed state preparation ($|\boldsymbol{s}_D^o < 1$)
forces (\ref{22})  to be satisfied as an inequality.

Finally, we analyze the increase of linear entropy in the SQDS.
Inserting (\ref{SQDS13}) into (\ref{26}) we have
\begin{equation}
\Delta \mathcal{G}_{Q}=\frac{1}{2}\mathcal{V}_{Q}^{o2}
                \left( 1-\mathcal{P}_D^{2}\right) \sin^2\Phi .
\label{SQDS16}
\end{equation}
Using Eq. (\ref{SQDS8}) and particularizing Eq. (\ref{24}) for the
Detecton, we rewrite Eq. (\ref{SQDS16}) as
\begin{equation}
\frac{2 \Delta \mathcal{G}_{Q}}{\mathcal{V}_{Q}^{o2}}=
    1-|\boldsymbol{s}_D^o|^2    + \mathcal{Q}_D^2,
\label{SQDS16bis}
\end{equation}
satisfying the general inequality given in (\ref{27}). As
commented below Eq. (\ref{27}), degradation of the purity of the
Quanton state is a necessary condition for the extraction of
quantum WWI ($\mathcal{Q}_D\neq 0$). However, it is not a
sufficient condition. This can be seen  from Eq. (\ref{SQDS16bis})
where $\Delta \mathcal{G}_{Q}$ can be nonvanishing for
$\mathcal{Q}_D=0$, provided the Detecton is prepared in a mixed
state ($|\boldsymbol{s}_D^o|< 1$).

In the case of pure state preparation of the Detecton, Eq.
(\ref{SQDS16bis}) simplifies to the expression
\begin{equation}
\frac{2\Delta \mathcal{G}_{Q}}{\mathcal{V}_Q^{o2}}
=\mathcal{Q}_{D}^2 . \label{SQDS17}
\end{equation}
Thus, in this case, the Quality of the detector directly gives the
increase in linear entropy of the Quanton, once normalized to its
initial visibility. The implications of (\ref{SQDS17}) are
straightforward. For $\mathcal{Q}_D=0$ there is no degradation of
the purity of the Quanton. For $\mathcal{Q}_D=1$, the linear
entropy increase to its maximum $\Delta
\mathcal{G}_{Q}=\frac{1}{2} \mathcal{V}_Q^{o2}$: maximum WWI is
extracted at the expense of the total degradation of the fringe
pattern of the Quanton, i.e., $\mathcal{V}_Q =0$, as demanded by
duality (see Eq. \ref{1.4bis}). Moreover, the Detecton
interferometer also degrades its visibility pattern. In order to
see this, we calculate $\mathcal{V}_D$ just by applying the
Q$\leftrightarrow$D symmetry to the labels of Eq. (\ref{SQDS13})
to obtain
\begin{equation}
\mathcal{V}_D = \mathcal{V}_D^o \,\sqrt{\cos^2 \Phi
+\mathcal{P}_Q^2 \sin^2\Phi } . \label{SQDS18}
\end{equation}
Combining Eq. (\ref{SQDS18}) and (\ref{SQDS13}), we obtain the
result
\begin{equation}
\left(  1-\mathcal{P}_Q^2 \right) \frac{\Delta
\mathcal{V}_Q^2}{\mathcal{V}_Q^{o2}} = \left(  1-\mathcal{P}_D^2
\right) \frac{\Delta \mathcal{V}_D^2}{\mathcal{V}_D^{o2}},
\label{SQDS19}
\end{equation}
where $\Delta \mathcal{V}_{\alpha}^2 \equiv \mathcal{V}_{\alpha}^2
- \mathcal{V}_{\alpha}^{o2}$, ($\alpha =\mathcal{Q},
\mathcal{D}$). Equation (\ref{SQDS19}) is valid for arbitrary
initial preparation of Quanton and Detecton.  In the case of
initial pure state preparation for both sub-systems, Eq.
(\ref{SQDS19}) reduces to the simple form
\begin{equation}
\Delta \mathcal{V}_Q^2 = \Delta \mathcal{V}_D^2.
 \label{SQDS20}
\end{equation}
The above equation provides a clear illustration on the reciprocal
effects of Duality in the SQDS. When WWI is extracted on one
system, both Quanton and Detecton degrade their interference
pattern by the same amount. In the more general case of mixed
state preparation, the balance in the reciprocal degradation is
weighted by the factors appearing in Eq. (\ref{SQDS19}). For
instance, consider the case in which we prepare the Quanton as the
total unpolarized state
$|\vec{s}_Q^o|=\mathcal{P}_Q=\mathcal{V}_Q^o=0$, and Detecton in a
pure state with $\mathcal{V}_D^o=1$. Since in the SQDS,
$\mathcal{Q}_D$ is independent on the state of the Quanton, full
WWI ($\mathcal{Q}_D=1$) can be extracted by the Detecton at the
expense of totally degrading its fringe visibility from
$\mathcal{V}_D^o=1$ to $\mathcal{V}_D=0$. When both systems are
prepared initially as a classical mixture, then
$\mathcal{Q}_D=\mathcal{Q}_Q=0$, and no WWI can be extracted no
matter the value of the entanglement between both interferometers.
The reciprocity between fringe degradation in Quanton and Detecton
Systems can be explained by the reciprocity between their
"quality" measures, since both of them are connected by duality.
In fact, applying the Q$\leftrightarrow$D symmetry to Eq.
(\ref{SQDS8}) we have
\begin{equation}
\frac{\mathcal{Q}_D}{\mathcal{V}_D^o} =
\frac{\mathcal{Q}_Q}{\mathcal{V}_Q^o} .
 \label{SQDS20bis}
\end{equation}
According to the above equation, any potential acquisition of WWI
is mutual. In other words, if the Detecton may acquired WWI about
the Quanton, the Quanton may acquired WWI about the Detecton. This
explains the simultaneous degradation of the fringe pattern of
both systems as one of them extracts WWI about the other.

The SQDS highlights the role of the initial visibility of both
interferometers in duality exchanges of WWI. As a matter of fact,
combining Eqs. (\ref{SQDS8}) and (\ref{SQDS13}), setting
$\mathcal{P}_D=\mathcal{P}_Q=0$, and applying the
Q$\leftrightarrow$D symmetry between labels, we obtain
\begin{equation}
\frac{\mathcal{D}_Q^2}{\mathcal{V}_D^{o2}} +
\frac{\mathcal{V}_Q^2}{\mathcal{V}_Q^{o2}} =
\frac{\mathcal{D}_D^2}{\mathcal{V}_Q^{o2}} +
\frac{\mathcal{V}_D^2}{\mathcal{V}_D^{o2}} = \cos^2 \Phi + \sin^2
\Phi =1 .
 \label{SQDS21}
\end{equation}
The above equation can be written as
\begin{equation}
\mathcal{D}_Q^2  + \frac{\mathcal{V}_D^{o2}}{\mathcal{V}_Q^{o2}}
\mathcal{V}_Q^2 = \mathcal{V}_D^{o2}.
 \label{SQDS21.bis}
\end{equation}
The duality implications of Eq. (\ref{SQDS21.bis}) are
straightforward. Since $\mathcal{V}_{\alpha}^2\leq
\mathcal{V}_{\alpha}^{o2}\leq 1$,$\mathcal{D}_Q =1$ implies both
$\mathcal{V}_D^o =1$ and $\mathcal{V}_Q =0$ as demanded by
duality. Maximum Detecton initial visibility is required to
totally degrade the Quanton's interference pattern. On the other
hand, $\mathcal{V}_Q =1$ in (\ref{SQDS21.bis}) forces
$\mathcal{D}_Q =0$. Thus, in contrast to the inequality given in
Eq. (\ref{1}), Eq. (\ref{SQDS21.bis}) is an equality. It is an
equality involving duality valid even for mixed state preparation
of Quanton and Detecton.

%Take now the Quanton prepared in a pure state and Detecton prepared in a mixed state. In order to illustrate Eq. (\ref{23}), we compute
%\begin{eqnarray}
%\label{SQDS16}
%&&\left( 1-\mathcal{P}_Q^2 \right)\mathcal{Q}_D^2 +\mathcal{P}_Q^2 +\mathcal{V}_Q^2
%\nonumber \\
%&=& \left( 1-\mathcal{P}_Q^2 \right)\left(\sin^2\Phi |\boldsymbol{s}_D^o|^2+\cos^2\Phi \right) +\mathcal{P}_Q^2  \le 1 ,
%\end{eqnarray}
%where we have used Eqs. (\ref{SQDS8}) and (\ref{SQDS13}), and taken into account that
%\begin{eqnarray}
%\label{SQDS17}
%|\boldsymbol{s}_D^o|^2 &=& \mathcal{P}_D^2 +  \mathcal{V}_D^2 \le 1 ,
%\nonumber\\
%|\boldsymbol{s}_Q^o|^2 &=& \mathcal{P}_Q^2 +  \mathcal{V}_Q^2 = 1 .
%\end{eqnarray}
%It can be seen from Eq. (\ref{SQDS16}) that the equality in (\ref{23}) is recovered for pure state preparation of the Detecton ($|\boldsymbol{s}_D^o|=1$).

%%%%%%%%%%%%%%%%%%%%%%%%%%%%%%%%%%%%%%%%%%%%%%%%%%%%%%%%%%%%%%%%%%%%%%%%%%
%%%%%%%%%%%%%%%%%%%%%%%%%%%%%%%%%%%%%%%%%%%%%%%%%%%%%%%%%%%%%%%%%%%%%%%%%%%%

\section{Summary}

We have introduced in this paper a quality measure $\mathcal{Q}$,
which characterizes how good a quantum detector is by quantifying
the maximum which-way information that it can acquire when placed
in a two-way interferometer. In this way, we are able to separate
the contribution to the distinguishability arising from the
quantum properties of the detector, given by~$\mathcal{Q}$, from
that stemming from the \textit{a~priori}  which-way knowledge
involved in the preparation of the interferometer. The latter is
given by the predictability~$\mathcal{P}$ characterizing how
asymmetrically the two-way interferometer is constructed.

In the spirit of \cite{Englert96}, we have derived an inequality relating $%
\mathcal{Q}$ and~$\mathcal{P}$ to the value of the fringe
visibility~$\mathcal{V}$ displayed at the output port of the
interferometer. This inequality allows us to quantify the
degradation of the fringe visibility~$\mathcal{V}$ involved in the
availability of the two kinds of which-way information. For
instance, it shows that maximum which-way information available
can be stored in the WWD, even for strongly asymmetric
interferometers. In addition we have shown that, in the case where
both systems are prepared in a pure state, both kinds of which-way
information represented by~$\mathcal{Q}$ and~$\mathcal{P}$ stand
on an equal footing concerning loss of coherence.

In the case of a classical WWD, $\mathcal{Q}$ can be regarded as
the quality of a noisy communication channel which is independent
of the state of the sender. Here, $\mathcal{Q}$ and~$\mathcal{P}$
are clearly separable as distinguishabilities stemming from
uncoupled sources of information; the total distinguishability is
just the maximum of the two. The situation turns out to be much
more complicated for a quantum WWD. Due to the entanglement
between Quanton and Detecton, the value of the quantum Quality
depends in general on the state of both systems.

We have applied our formalism to a  quantum logic gate: the
Symmetric Quantum-Detecton system (SQDS). This system can be
regarded as two coupled interferometers trying to acquire WWI
about each other. In the SQDS both interferometers are coupled by
a dispersive interaction. There is no energy transfer between
Quanton and Detecton, just an information transfer established via
quantum correlations. We have shown that in the SQDS the
inequality involving duality in terms of the distinguishabilities
$\mathcal{Q}$ and $\mathcal{P}$ is more stringent than the
Englert큦 inequality given in terms of the total
distinguishability $\mathcal{D}$. Also, our formalism has shown
useful in order to characterize the reciprocal effects induced by
duality on both systems. Finally, we have shown an equality
involving duality for the SQDS. This equality highlights the role
of the initial visibility of both systems as limiting factors in
the mutual transfer of WWI between both qubits.

%
%%%%%%%%%%%%%%%%%%%%%%%%%%%%%%%%%%%%%%%%%%%%%%%%%%%%%%%%%%%%%%%%%%%%%%%%%
%%%%%%%%%%%%%%%%%%%%%%%%%%%%%%%%%%%%%%%%%%%%%%%%%%%%%%%%%%%%%%%%%%%%%%%%%
%

\begin{acknowledgments}
We thank B.-G. Englert for his constructive comments. The authors are also
grateful to the Max-Planck-Gesellschaft (Quantenoptik). J. M.-L. was
initially supported by the TMR program of the European Union (Marie Curie
fellowship) under contract No.\ ERBFMBICT972392 and then by CIC from
Universidad Michoacana de San Nicol\'{a}s de Hidalgo, and the PROMEP\
program in M\'{e}xico. D.A.H. obtained support in the earlier stages of this
work from the Office of Basic Energy Sciences, U.S. Department of Energy.
\end{acknowledgments}

\appendix*
\section{}
In this appendix, we show that the quantity
\begin{equation}
f_Q \equiv
\frac{(1-\mathcal{P}_Q^2)(1-\mathcal{Q}_D^2)}{(1-\mathcal{D}_Q^2)}
, \label{SQDSf2}
\end{equation}
satisfies $0\leq f_Q\leq 1$. Since $\mathcal{D}_Q = \text{Max}
\left\{ \mathcal{P}_Q, \mathcal{R}_Q \right\}$, we consider two
cases separately. First, take $\mathcal{P}_Q \geq \mathcal{R}_Q$.
Here $\mathcal{D}_Q=\mathcal{P}_Q$ and
$f_Q=(1-\mathcal{Q}_D^2)\leq 1$. Else, for $\mathcal{P}_Q <
\mathcal{R}_Q$ we have
\begin{equation}
f_Q = \frac{(1-\mathcal{P}_Q^2)(1-\mathcal{Q}_D^2)}{
1-\mathcal{P}_Q^2 |\boldsymbol{s}_D^o|^2 - \mathcal{Q}_D^2
 ( 1-\mathcal{P}_Q^2)} ,
\label{SQDSf3}
\end{equation}
where we have used Eq. (\ref{SQDS11rewritten}). Let us define the
auxiliary quantities
\begin{eqnarray}
\xi &\equiv& (1-\mathcal{P}_Q^2)(1-\mathcal{Q}_D^2) \geq 0,
\nonumber \\
g&\equiv& 1-\mathcal{R}_Q^2 -\xi = 2\mathcal{P}_Q^2\mathcal{Q}_D^2
+\mathcal{P}_Q^2 ( 1- |\boldsymbol{s}_D^o|^2) \geq 0 .
\label{SQDSf4}
\end{eqnarray}
In terms of $g$, Eq. (\ref{SQDSf2}) can be written as
\begin{equation}
f_Q = \frac{\xi}{g+ \xi } . \label{SQDSf5}
\end{equation}
Therefore, since $g\geq 0$ we conclude $0\leq f_Q \leq 1$.

%\bibliography{Quality}               % Produces the bibliography via BibTeX.

\begin{thebibliography}{19}
\bibitem[{eps()}]{epsilon}
\bibinfo{note}{For the sake of simplicity, we have taken a symmetric channel
  characterized by a single probability error~$\epsilon $. Note that this is
  independent of the asymmetry of the quanton, characterized by $ w_{+}\neq
  w_{-}$.}
\bibitem[{lab()}]{label}
\bibinfo{note}{${\cal D}$ in (\ref{SQDS10}) is the distinguishability of the
  ways of the Quanton interferometer. Hence the label ${\cal D}_Q$. On the
  other hand, the Quality ${\cal Q}_{\alpha}$ (${\cal Q}_Q, {\cal Q}_D$) of
  each interferometer measures their properties as detectors of WWI about the
  other interferometer. Thus, according to our notation, ${\cal Q}_D$
  contributes to ${\cal D}_Q$}.
\end{thebibliography}

\end{document}